\documentclass[10pt,conference]{IEEEtran}
\usepackage{amsmath,amssymb,amsfonts}
\usepackage{algorithmic}
\usepackage{graphicx}
\usepackage{textcomp}
\usepackage{comment}
\usepackage[hyphens]{url}

\def\BibTeX{{\rm B\kern-.05em{\sc i\kern-.025em b}\kern-.08em
    T\kern-.1667em\lower.7ex\hbox{E}\kern-.125emX}}

\usepackage{mathptmx} 

\usepackage{fancyhdr}
\usepackage[normalem]{ulem}
\usepackage[hyphens]{url}
\usepackage[sort,nocompress]{cite}

\usepackage[final]{microtype}
\usepackage[bookmarks=true,breaklinks=true,letterpaper=true,colorlinks,citecolor=blue,linkcolor=blue,urlcolor=blue]{hyperref}

\usepackage{subfig}
\usepackage[table]{xcolor}
\usepackage{wrapfig}
\usepackage{arydshln}
\usepackage{multirow}
\usepackage{soul}
\usepackage{color}
\usepackage{balance}

\DeclareRobustCommand{\hlpink}[1]{{\sethlcolor{pink}\hl{#1}}}
\mathchardef\mhyphen="2D

\def\BibTeX{{\rm B\kern-.05em{\sc i\kern-.025em b}\kern-.08em
    T\kern-.1667em\lower.7ex\hbox{E}\kern-.125emX}}

\newcommand{\nej}[1]{{{\color{red}NEJ: {#1}}}}
\newcommand{\rc}[1]{{{\color{magenta}From Robert: {#1}}}}
\newcommand{\djs}[1]{{{\color{red}DJS: {#1}}}}
\newcommand{\bdtext}[1]{{{\color{blue}{#1}}}}

\makeatletter
    \newcommand{\linebreakand}{%
      \end{@IEEEauthorhalign}
      \hfill\mbox{}\par
      \mbox{}\hfill\begin{@IEEEauthorhalign}
    }
    \makeatother

\usepackage[font=small]{caption}

\title{Low-Energy Line Codes for On-Chip Networks}

\author{
\IEEEauthorblockN{Beyza Dabak}
    \IEEEauthorblockA{\textit{Duke University}\\
    beyza.dabak@duke.edu}
    \and
    \IEEEauthorblockN{Major Glenn}
    \IEEEauthorblockA{\textit{Duke University}\\
    major.glenn@duke.edu}
    \and
\IEEEauthorblockN{Jingyang Liu}
    \IEEEauthorblockA{\textit{University of Toronto}\\
    leonjy2014@gmail.com}

\and
\IEEEauthorblockN{Alexander Buck}
    \IEEEauthorblockA{\textit{University of Toronto}\\
    alexander.buck@mail.utoronto.ca} 
\linebreakand
\IEEEauthorblockN{Siyi Yang}
    \IEEEauthorblockA{\textit{Duke University}\\
    siyi.yang@duke.edu}
\and

\IEEEauthorblockN{Robert Calderbank}
    \IEEEauthorblockA{\textit{Duke University}\\
    robert.calderbank@duke.edu}
\and

\IEEEauthorblockN{Natalie Enright Jerger}
    \IEEEauthorblockA{\textit{University of Toronto}\\
    enright@ece.utoronto.ca}
\and

\IEEEauthorblockN{Daniel J. Sorin}
    \IEEEauthorblockA{\textit{Duke University}\\
    sorin@ee.duke.edu}
\and
\IEEEauthorblockN{}
}

\begin{document}
\maketitle
\thispagestyle{plain}
\pagestyle{plain}



\begin{abstract}

Energy is a primary constraint in processor design, and much of that energy is consumed in on-chip communication.  Communication can be intra-core (e.g., from a register file to an ALU) or inter-core (e.g., over the on-chip network).  In this paper, we use the on-chip network (OCN) as a case study for saving on-chip communication energy.
We have identified a new way to reduce the OCN’s link energy consumption by using line coding, a longstanding technique in information theory. Our line codes, called Low-Energy Line Codes (LELCs), reduce energy by reducing the frequency of voltage transitions of the links, and they achieve a range of energy/performance trade-offs. 



\end{abstract}

\section{Introduction}
\label{sec:intro}

Architects strive to save energy consumption wherever they can, and within the chip, one of the largest culprits is on-chip communication~\cite{dally-keynote}. Communication occurs both within a core (e.g., from register file to L1D cache, on pipeline bypass paths, etc.) and between cores via the on-chip network (OCN).
To reduce on-chip communication energy consumption, we exploit and adapt ideas from information theory. Information theory explores how best to send information over communication channels, and it is applicable to the communication channels on processor chips.  In particular, we use the information theory technique of \textit{line coding} and apply it to processor architecture.

Line coding, which consists of two interrelated aspects---\textit{modulation} and \textit{data coding}---is a longstanding technique for transmitting data over a communication channel (i.e., wire).  Modulation is how we represent 0s and 1s with voltages, such as with the widely used NRZ (non-return-to-zero) and NRZI (non-return-to-zero inverted) signaling techniques. With NRZ signaling, a logical 0 is mapped to low voltage and a logical 1 is mapped to high voltage. With NRZI  signaling, a 0 is mapped to an unchanged voltage, and a 1 is mapped to a voltage transition.  
Coding is how we map the $k$-bit dataword into the $n$-bit codeword, so as to achieve desirable transmission properties.
As a simple example, a parity code adds a single bit to the end of the dataword to create a codeword that can detect single-bit errors. Other codes create codewords with other properties, such as RLL codes which limit the length of a run of like symbols in any codeword~\cite{tang:iandc:1970}.  Code constraints enable us to shape the distribution of 0s and 1s in the codewords.

\begin{figure}[t]
    \centering
    \subfloat{
    \includegraphics[trim={1.25in 0in 1.5in 0in},width=0.41\textwidth]{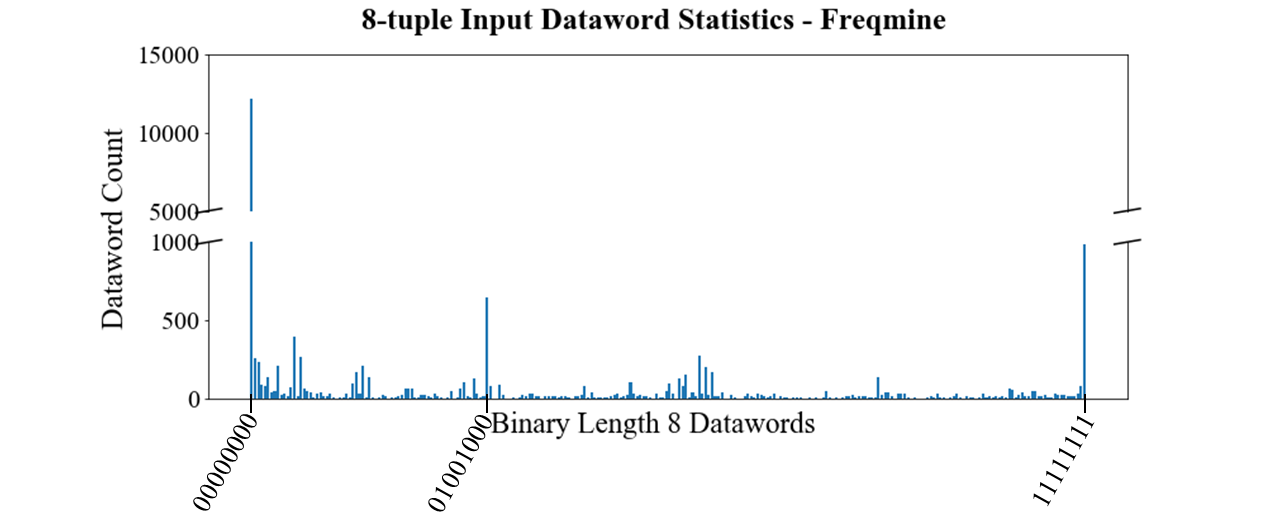}\vspace{0.05in}} \\
    \subfloat{
    \includegraphics[trim={1.25in 0in 1.5in 0in},width=0.41\textwidth]{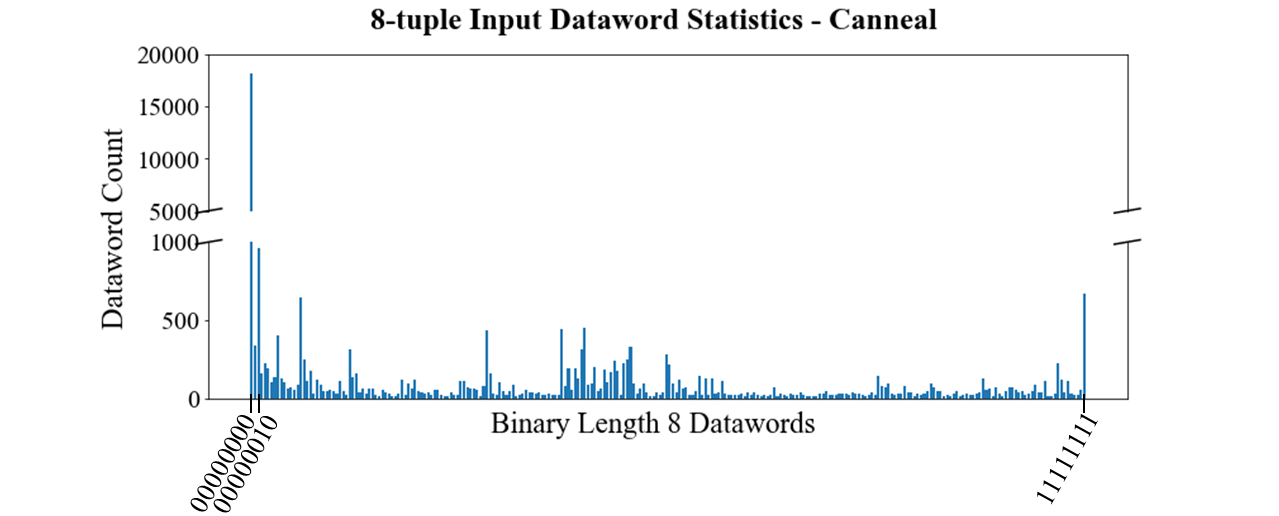}\vspace{0.05in}} \\
\subfloat{    
\includegraphics[trim={1in 0in 1in 0in},width=0.34\textwidth]{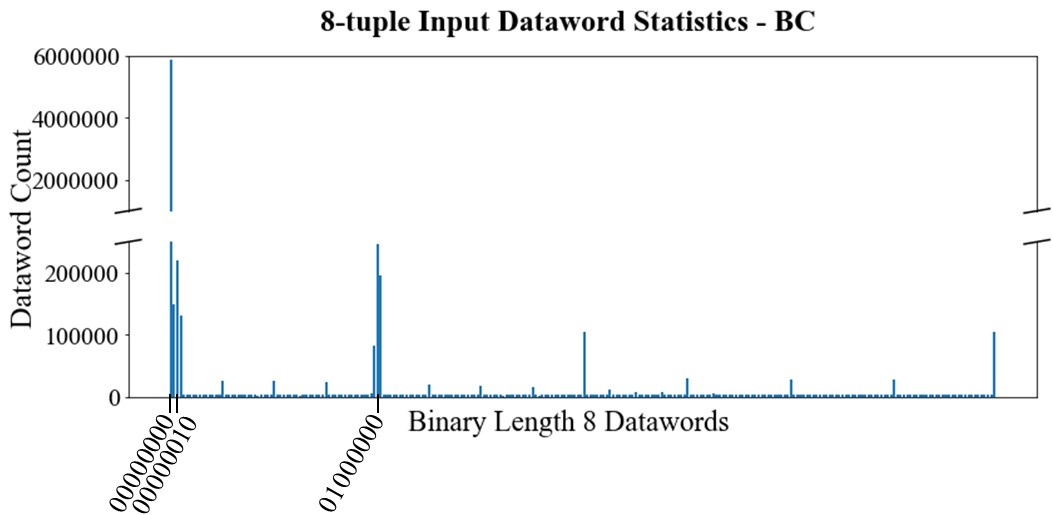}\vspace{0.0in}}       
\caption{Dataword distributions for benchmarks (2 CPU and 1 GPU)}
    \label{fig:blackscholes-dist}
    \vspace{-0.1in}
\end{figure}

To reduce energy in on-chip communication, we introduce line codes that we call Low-Energy Line Codes (LELCs). 
Energy is a function of the number of voltage transitions; therefore, we seek to reduce transitions. We use NRZI signaling; thus, the goal is to reduce the number of 1s.  
Our LELCs take advantage of the observation that communicated data communicated on chip is not uniformly random (i.e., the $k$-bit datawords are not equiprobable), as assumed in much prior work.  Fig.~\ref{fig:blackscholes-dist} presents the distribution of 8-bit datawords sent through the OCN for two representative CPU benchmarks and one GPU benchmark.\footnote{For full list of workloads and methodology, see Section~\ref{sec:methodology}.} 
The all-0 dataword is by far the most common, the all-1 dataword is fairly common, and other datawords are not uniformly likely.\footnote{Other benchmarks had similar trends of all-0 and all-1 datawords, along with variations in the next most common set of datawords.}
These results are consistent with intuition and prior experimental results, including results that show that all-0 datawords are very common~\cite{lipasti:asplos:2007}.
We design codes that exploit redundancy in the data to produce codewords with fewer 1s.

While many data codes exist that enable us to reduce the number of 1s, we must consider their costs.   Specifically, the impact of coding on performance is represented by its \textit{rate}, which is the ratio of dataword bits to codeword bits.  Parity, for example, requires us to send $n=k+1$ codeword bits and thus has a rate of $k/(k+1)$.  
Information theory governs the pareto optimal trade-offs between energy reduction and rate.  However, just because a trade-off is theoretically possible does not imply that it is practical; certain codes might require prohibitively complex or large hardware for encoding and decoding. 
We also must ensure that any line code we develop does not exacerbate crosstalk among the wires in a link. 

In this paper, we present several practically implementable line codes that correspond to a range of energy-rate trade-offs.  
To showcase the opportunity for LELCs, we focus on OCN communication; the OCN takes up approximately 10\% of the area~\cite{daya2014scorpio} and consumes up to 20\% of the processor's power budget~\cite{howard10,daya2014scorpio,Hoskote2007mesh, sewell:swizzle:2012}.
Furthermore, OCN links contribute a substantial fraction of OCN power--$\sim$50\% of OCN dynamic power is spent on wires~\cite{onur-asplos}. Transmitting a {\it single bit} on a link consumes approximately $1-1.75$pJ~\cite{sun:nocs:2012, hotchips-2019}.
In the context of the OCN, having a range of trade-offs is attractive, because of the over-provisioning of link bandwidth and the varying runtime demand for bandwidth.  Furthermore, we provide a mechanism for throttling coding when OCN utilization is high.
This paper makes the following primary contributions:
\begin{itemize}
    \item \textls[-15]{Introduces line codes to reduce OCN link energy.}
    \item Presents insight into how OCN architecture influences which codes are effective and which are less so.
    \item Implements low-cost hardware implementations of encoders and decoders for the most promising LELCs.
    \item Provides a simple dynamic strategy to turn off line coding when bandwidth demand is high.
    \item Experimentally shows that our LELCs provide a tunable range of energy-performance trade-offs with energy reductions ($8.0\%\mhyphen36.7\%$) that greatly exceed runtime increases ($-1.25\%\mhyphen8.42\%$).  Our LELCs also reduce crosstalk by an average of 21.3-36.1\%.
    \item Demonstrates that codes designed for CPU workloads provide significant energy improvements in GPUs.
\end{itemize}
\section{System Model}
\label{sec:system-model}

We focus on the links in an OCN, and we are largely agnostic to the routers and topology.  Coding happens at the source and destination, leaving routers unchanged. We assume a 2D mesh but our insights generalize to other topologies.\footnote{We focus here on OCNs for CPUs. In Sec.~\ref{sec:discussion}, we present link energy reductions for GPUs which typically feature crossbar networks.} Link energy is proportional to link length; this Manhattan distance between cores is independent of the number of router traversals between them (router traversals would be a function of topology).


\noindent{\bf OCN Traffic.} OCN traffic includes two types of packets: control and data.  A control packet (e.g., coherence request or acknowledgment) is short (e.g., $8$B), because it is effectively just a header. A data packet (e.g., coherence response) is long because it carries a cache block of data (e.g., $64$B) in addition to a header.  In this paper, we only perform coding on data packets, because of the greater opportunity. 
Each data packet is divided into multiple flits whose size corresponds to the width of the link. 
Furthermore, we only encode the data payload and not the header, because coding the header would introduce routing complexity and latency (i.e., having to decode the header at each router to determine the next hop). 

Our OCN has fixed link widths between routers, and we must map the bits being transmitted to those fixed number of wires. Coding will change the rate; a change in rate can be accounted for by increasing the number of flits per packet and/or by increasing the link width. We conservatively assume that we cannot widen links. For example, if our $64$B data payloads are broken up into 4 flits of $16$B each and we reduce the rate to $0.9$, we would require one additional flit to accommodate the coded data.  Additional flits consume bandwidth on the links and in the router crossbars and also increase serialization latency (paid only once at the destination).  

The increase in the number of flits is discretized because of the fixed link width.  Returning to our example with a link width of $16$B, any increase in data payload size from 1 bit to $16$B requires exactly one flit to accommodate it.  Thus, a code that adds $16$B has the same performance impact as one that adds anything less, and thus it often makes sense to use codes that add multiples (or near multiples) of the link width.


\noindent{\bf Energy Consumption.}
Recent work demonstrates that links consume $\sim40\mhyphen50\%$ of the total energy in the network~\cite{kar2017case, werner2017designing,onur-asplos}. 
In the context of the un-core (including L1, L2, OCN and DRAM), prior work reports that links consume an average of $\sim30\%$ of uncore  energy~\cite{kurian2014locality}. 
Link energy is dominated by the energy consumed when the voltage on wires transitions (from low-to-high or high-to-low).  The amount of energy consumed during a transition depends on many factors--wire length, resistivity, and parasitic capacitance and inductance--in this work, we focus on reducing the number of transitions as a proxy for the energy savings. 


\noindent{\bf Link Geometry.}
A link is a collection of wires, and these wires are arranged in some 3D geometry.  For the purposes of this work, link geometry matters only insofar as it affects crosstalk (discussed next).  
We assume the wires in a link are arranged such that a cross-section of them is a 2D grid. 


\noindent{\bf Crosstalk.}
Crosstalk is the phenomenon in which a signal transmitted through one information channel affects signals on neighboring channels. Crosstalk results from capacitive or inductive coupling between adjacent parallel wires, and transmission errors are more likely to occur when the interference from neighboring wires is constructive. This happens when the neighbors of a victim wire experience identical transitions and the victim wire experiences no transition or a transition in the opposite direction. Crosstalk manifests as a delay proportional to the effective capacitance experienced by the victim wire, and it is a significant issue in on-chip links~\cite{duan:hot-interconnects:2001, sridhara:tcad:2007}.

\section{The Potential of Line Coding}
\label{sec:limit-study}

The principles of information theory enable us to analytically bound the potential of line coding.  To do so, we must make an assumption about the input datawords; specifically, we assume they are equiprobable. Real-world inputs are not equiprobable, but with lossless compression, if the input is compressed down to its entropy, it can be made equiprobable. 



First, consider rate.  The maximum achievable rate $R$ of a code that has a probability $f$ that any given bit in the codeword is a 1 
is determined by the binary entropy function $H(f)$:
\begin{align}
R = H(f) = -flog_{2}f - (1-f)log_{2}(1-f).     
\end{align}
Intuitively, the rate is 0 at the extreme values of $f$; if we send only 0s or only 1s, we convey no information. The maximum achievable rate of 1 occurs when $f=\frac{1}{2}$, i.e., when 0s and 1s are equally likely, as with uncoded data.\footnote{The converse is not true; one can create codes with $f=\frac{1}{2}$ that have rate less than the upper bound of 1.}

\begin{figure}[t]
    \centering
    \includegraphics[clip=true,trim=0in 0in 0in 0.37in, width=0.32\textwidth]{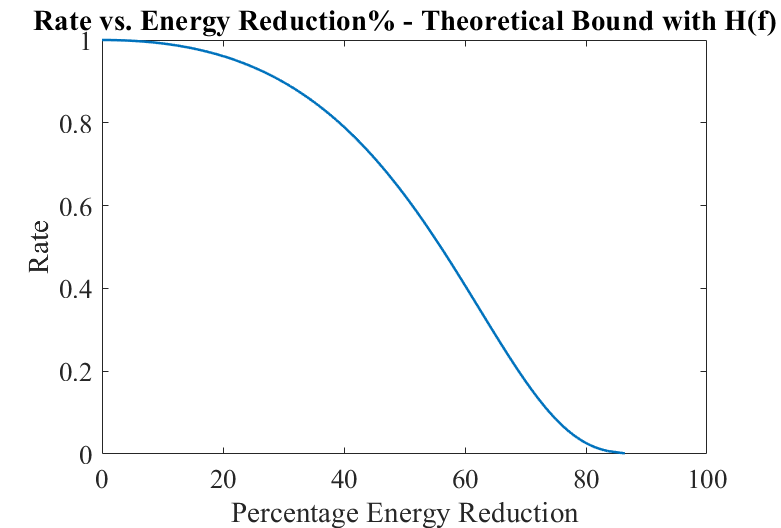}
    \caption{Optimal trade-off between rate and energy reduction for equiprobable input data. }
    \label{fig:rate-vs-f}
    \vspace{-0.2in}
\end{figure}

Our goal in this work is to save energy (i.e., decrease $f$) while sacrificing as little rate as possible.  
We plot rate as a function of the percentage of energy reduction in Figure~\ref{fig:rate-vs-f}.  
The figure shows the trade-offs that are possible with line coding, with two caveats. First, recall that this assumes equiprobable input datawords, which is not true of real application data though it could be achieved via compression. 
Second, some of the points on the curve may correspond to codes that would be prohibitively expensive to implement, due to circuit latency and/or energy.  Among the possible trade-offs, we believe that there is a sweet spot for OCNs at a rate around $0.8$, which has a corresponding theoretical energy reduction of $40\%$. 


It is natural to consider compression as an alternative to line coding because our benchmarks are very compressible, and compression reduces the number of bits communicated. Consider communicating an $N$ bit file that can be compressed to $NR$ bits, where the information rate $R<1$. If we assume perfect compression (down to the information rate $R$), then bits in the compressed file will be equiprobable, and so we will transmit $NR/2$ 1s. In a world without practical constraints, we can design a line code comprising $2^{RN}$ codewords of length $N$, each of which contains at most $H^{-1}(R)N$ 1s, where $H$ is the binary entropy function. When the information rate $R=1/10$ (typical of our benchmarks) we have $NR/2 = (0.05)N$ versus $H^{-1}(R)N \approx (0.013)N$, so compression requires more energy to send the file. It is important to note that this ideal analysis assumes perfect compression and perfect coding, and it neglects the practical considerations of implementing either compression or coding on-chip: latency and area.

\section{Low-Energy Line Codes for OCNs}
\label{sec:newcodes}

Our goal is to reduce OCN energy by using NRZI signaling and by reducing the frequency of 1s  transmitted. There are several types of codes that we could use, adapt, or create to shape the frequency of 1s, and we exploit the rich information theory tradition of coding for other contexts and applications.
We call these codes Low-Energy Line Codes (LELCs). First, we  outline the properties we seek when developing our codes: 

\begin{itemize}
    \item Reduce the fraction of 1s to a sufficient extent to achieve substantial energy savings, even when accounting for additional overheads such as encoding/decoding circuitry.
    \item Achieve a rate that is high enough to incur a penalty of only one additional flit.  This rate threshold depends on the link width of the target OCN system (e.g., it is $0.8$ in our experimental system).
    \item Facilitate low-latency hardware for encoding and decoding, as well as the ability to easily parallelize the hardware for encoding (decoding) independent datawords (codewords).  
    \item \textls[-5]{Generalize well to different benchmarks and systems; per application codes are possible but would require considerable profiling and reconfigurability to achieve good results.}
    \item Avoid pathological scenarios that could increase energy. 
    \item Reduce crosstalk as a result of reducing transitions; in the worst case, avoid increasing crosstalk.
\end{itemize}

\subsection{LELC Class 1: Flip-N-Write}

A simple way to reduce the frequency of 1s in the codeword stream is by inverting datawords with more 1s than 0s.  The cost is an extra flag bit that denotes whether the codeword's bits were flipped or not; we refer to this extra bit as the IsFlipped bit.  Thus, for $k$-bit datawords, we have $k+1$-bit codewords, resulting  in a rate of $k/(k+1)$. 
This code is a degenerate case of coset coding~\cite{forney:ieeetransit:1988a, forney:ieeetransit:1988b}. The code is parameterized by $k$, and it has the nice property that a codeword will never have more than half of its bits equal to 1.
This coding scheme has been previously proposed for use in storage, under the name Flip-N-Write~\cite{cho:micro:2019}, which we adopt.  It has also been used for general data buses, under the name bus-invert~\cite{stan:1995}.   While bus-invert is also an application of the code for communication, that work did not consider real-world data or any specific network, but instead assumed equiprobable datawords.

\begin{figure}[t]
    
    \centering
    \includegraphics[trim=0 0.5in 0 1in, width=0.42\textwidth]{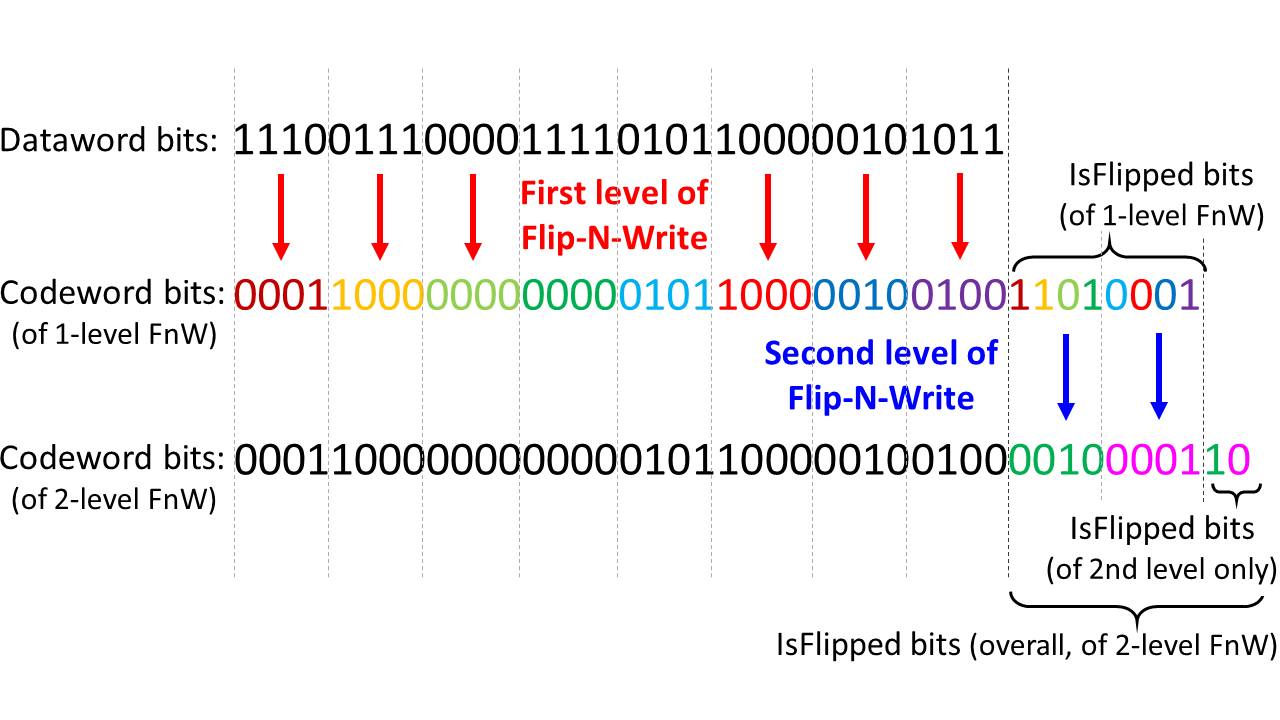}
    \caption{Illustration of 2-level Flip-N-Write (k=4).} 
    \label{fig:2levelFnW}
    \vspace{-0.2in}
\end{figure}

We extend Flip-N-Write (FnW) with a multi-level version that further trades rate for energy savings.  
Consider applying (single-level) FnW on a stream of datawords.  The result is a stream of codewords, each with its own IsFlipped bit. 
Now, if we consider each group of $f$ IsFlipped bits (i.e., the IsFlipped bits corresponding to $f$ codewords, where $f$ may or may not equal $k$), we could treat them as a $f$-bit dataword to be encoded with FnW.  Thus, for every group of $f$ datawords, we have $f$ codewords, each with $k+1$ bits, plus one extra bit to denote whether the $f$ IsFlipped bits are inverted or not.  We illustrate an example in Figure~\ref{fig:2levelFnW}.
FnW is attractive in that encoding and decoding require only simple logic, thus adding minimal complexity, latency, and energy consumption.  We evaluate the circuitry and energy for encoding and decoding FnW (and all other LELCs) in our experimental evaluation (Sec.~\ref{sec:evaluation}).

\subsection{LELC Class 2: Tree Codes}

\begin{figure}[t]
    \centering
    \includegraphics[trim={0in 0.9in 0in 0.0in},width=0.28\textwidth]{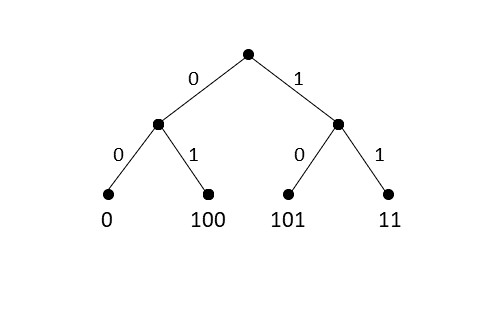}
    \caption{Huffman coding tree with codewords at leaves. Assumes the frequency of input datawords is, in decreasing order: 00, 11, 01, and 10. These datawords are mapped to codewords 0, 11, 100, and 101.}
    \label{fig:huffman}
    \vspace{-0.2in}
\end{figure}

One can create a code using a tree, in which the leaves of the tree are the codewords, and the path taken to reach a leaf is the dataword.  
A famous example of a tree code is a Huffman code~\cite{huffman:pire:1952}.  
A common use of a Huffman code is lossless compression, and it achieves compression by assigning the shortest codewords to the most common datawords.  
Figure~\ref{fig:huffman} shows a simplified example which maps $2$-bit data words to code words that are $1$ to $3$ bits in length.\footnote{\textls[-10]{In lossless compression, a Huffman code encodes source strings into variable-length codewords. Source symbols label leaves of the Huffman tree, and codewords label paths to the root.  To be consistent with our other tree codes, we reverse this assignment, labeling leaves by codewords and paths by datawords.}} 
In this example, input 00 is most common so it is mapped to the shortest (single-bit) codeword. Uncommon datawords are mapped to longer codewords. 
Because a Huffman tree is designed to compress datawords with expected (e.g., profiled) input statistics, its rate is greater than 1 when input statistics are similar to expected.

\begin{figure}[t]
    \centering
    \includegraphics[width=0.33\textwidth]{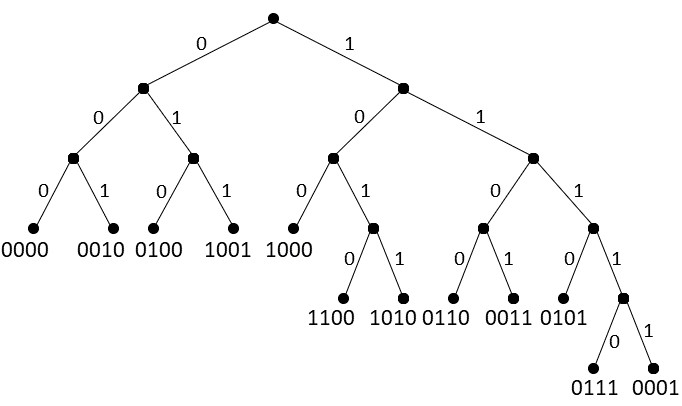}
    \caption{Tree Code 1 (TC1): Variable rate that is bounded between $3/4$ and $5/4$. 1 bit added redundancy for all 0s and compression opportunity for strings of 1s.} 
    \label{fig:tree3}
    \vspace{-0.1in}
\end{figure}

\begin{figure}[t]
    \centering
    \includegraphics[width=0.33\textwidth]{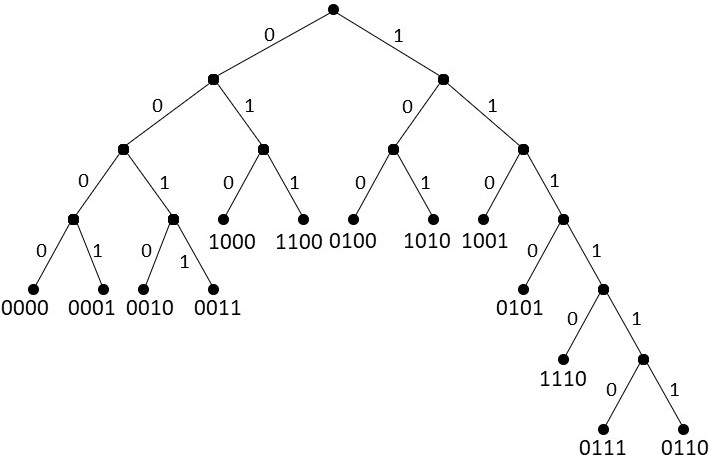}
    \caption{Tree Code 2 (TC2): Variable rate that is bounded between $3/4$ and $6/4$. No added redundancy for all 0s and compression opportunity for strings of 1s.}
    \label{fig:tree2}
    \vspace{-0.1in}
\end{figure}

One can also construct tree codes that do not follow the Huffman algorithm.  Instead of compression, our primary goal is to reduce the frequency of 1s in the codewords; nevertheless, any compression we achieve benefits our code rate.
Therefore, we construct trees that map dataword bit sequences to codeword sequences with fewer 1s.  While doing so, it is often possible to retain some of the compression benefits of Huffman codes.  
The construction of a tree code is informed by the expected dataword distribution from PARSEC benchmarks. 
Recall from Figure~\ref{fig:blackscholes-dist}, the distribution of datawords is not equiprobable.
We have developed several tree codes that provide different energy/rate trade-offs and require different amounts of hardware for encoding/decoding.  In Figures~\ref{fig:tree3} and~\ref{fig:tree2}, we show two of the tree codes we have developed. Both tree codes guarantee that every codeword has no more 1s than its corresponding dataword, and both target compression of strings of dataword 1s (i.e., the rightmost parts of the trees).  Compared to tree code 1 (TC1), TC2 targets longer strings of dataword 1s. The essential design point in TC2 is that no bit redundancy is added to the all-zero dataword, which is the most common dataword in our benchmarks and very common, in general~\cite{lipasti:asplos:2007}. Thus, TC2 offers higher rate for the most frequent datawords and a higher overall rate, but less energy savings than TC1. 





We designed TC1 and TC2 to fairly compete with FnW$_{k=3}$, 
such that the codeword length $n=4$ in both schemes, and the lower bound on rate for TC1 and TC2 is $3/4$. (We can make a tree code that is $exactly$ equivalent to FnW$_{k=3}$ with a balanced tree that has path length 3 and codeword length $n=4$.\footnote{In fact, the balanced left part of TC1 is equivalent to FnW$_{k=3}$.}) 
A small value of $k$ limits the depth of the tree and the logic required to encode the datawords. 
Using a similar methodology, we could design other tree codes to compete against FnW with larger values of $k$. As we show in Section~\ref{sec:circuits}, though, even a shallow tree has a high encoding latency due to variable length encodings; therefore, we do not consider deeper trees further in this paper.



Tree codes can be either fixed or variable rate (both TC1 and TC2 are variable rate), depending on whether the tree is perfectly balanced or not, respectively.  Fixed-rate codes generally enable simpler hardware for encoding/decoding, but 
variable-rate codes are attractive when they offer better energy/rate trade-offs.  We discuss how to overcome the implementation challenges of variable-rate codes in Section~\ref{sec:circuits}.

\subsection{LELC Class 3: Mapping Codes}

Fundamentally, a code is a mapping from a dataword to a codeword.  Mappings are often based on math or trees that facilitate simple hardware.  However, for short codes, we can simply create a 1-to-1 mapping using a lookup table.  

We take advantage of this opportunity by explicitly mapping the most common $k$-bit datawords to the $n$-bit codewords with the lowest $weight$, i.e., number of 1s.  The rate of a mapping code is under our control, because we can choose $n$ to be equal to $k$ (in which case rate equals $1$) or greater.  As $n$ increases, rate decreases, but we can reduce energy more.  Consider datawords of length $k$ and codewords of length $n=k+1$, and assume $k$ is even, for simplicity. Datawords can have a weight anywhere from $0$ to $k$.  For codewords, we can choose to use the $2^k$ $n$-bit strings with weights from $0$ to $k/2$. We will show later that mapping codes with length $k+1$  reduce energy significantly more than mapping codes with length $k$, at a relatively small loss in rate (i.e., rate $k/(k+1)$ versus rate $1$).
The primary constraint on the size of a mapping code is $k$, because the mapping table has $2^k$ entries, each of which is $n$ bits long. Large tables may be too slow and energy-hungry to be viable. 

In Table~\ref{table:mapping_code1}, we illustrate a simple mapping code example with $3$-bit datawords and $3$-bit codewords; because datawords and codewords are the same length, the code rate equals $1$.  Datawords are sorted in descending order according to observed frequency in data packets, and the corresponding codewords are in ascending order according to their weights. 

\begin{table}[tb]
\centering
\caption{Mapping Codes}
\subfloat[Example rate-$1$]{
\centering
{\small
\begin{tabular}{ | c | c | }
    \hline
    Dataword & Codeword  \\ \hline
    000 & 000 \\ \hline
    111 & 001 \\ \hline
    100 & 010 \\ \hline
    001 & 100 \\ \hline
    010 & 011 \\ \hline
    110 & 101 \\ \hline
    011 & 110 \\ \hline
    101 & 111 \\ \hline
\end{tabular}
\label{table:mapping_code1}
}
}
\hspace{0.1in}
\subfloat[Example rate-$3/4$]{
\centering
\small{
\begin{tabular}{ | c | c | }
    \hline
    Dataword & Codeword  \\ \hline
    000 & 0000 \\ \hline
    111 & 0001 \\ \hline
    100 & 0010 \\ \hline
    001 & 0100 \\ \hline
    110 & 0011 \\ \hline
    010 & 1000 \\ \hline
    011 & 0101 \\ \hline
    101 & 0110 \\ \hline
\end{tabular}

\label{table:mapping_code2}
}
}
\vspace{-0.3in}
\label{table:mapping_code}
\end{table}

Mapping codes offer the possibility of guaranteeing that $every$ codeword has a weight that is no greater than its corresponding dataword.  This feature could be desirable for making guarantees about energy consumption, as well as serving as a safety net against adversarial input data. A rate-$1$ mapping code cannot provide this guarantee, but once rate is $k/(k+1)$ or less, we can create maps with this guarantee. Such maps offer better worst-case behavior at the potential cost of somewhat worse common-case behavior.  We show a rate-$3/4$ mapping code with this property in Table~\ref{table:mapping_code2}. Comparing the examples in Table~\ref{table:mapping_code}, note that dataword {\it 101} is mapped to codeword {\it 111} which increases its energy when $n=3$ (rate-$1$) but is mapped to {\it 0110} when $n=4$ (rate-$3/4$), which provides the aforementioned guarantee on energy consumption. 


The main challenge in designing a mapping code is choosing the mapping. 
As with our tree codes, we use profiling similar to Fig.~\ref{fig:blackscholes-dist}.
We profile PARSEC benchmarks~\cite{parsec} independently and in aggregate, and we explore how well per-benchmark mappings perform on other benchmarks (i.e., how well does the mapping based on the profile for benchmark X perform on benchmark Y) and how well the aggregate mapping performs on each benchmark.  The similarities between benchmark distributions enables a single aggregate mapping to work well; we quantify these results in Sec.~\ref{sec:evaluation}.
Note that, while FnW targets high-weight datawords, regardless of their observed frequency, mapping codes (like tree codes, to a somewhat lesser extent) exploit differences in observed frequencies. Mapping codes also have the potential to be reconfigurable; we leave the study of this to future work.


\subsection{Compound LELCs}


One can combine multiple codes to achieve multiple goals, and we refer to such codes as \textit{compound codes}.  Many compound codes are possible, and we now discuss one compound LELC which combines a simple compression code with a code that is designed to save energy.\footnote{\textls[-10]{Recall from Sec.~\ref{sec:limit-study} that compression alone is less effective than line coding.}}  
The dataword is first converted by the compression scheme (discussed next) into an intermediate codeword. At this step the objective is increasing the code rate. Then the intermediate codeword is converted by the rate-$8/9$ mapping code into the final codeword. At this step the objective is energy reduction.

Our compression scheme exploits the observation that long runs of 0s are very prevalent in typical software (including our benchmarks). 
Our compression scheme is as follows: a length-$k$ dataword consisting of all 0s is mapped to `1' and all other length-$k$ datawords are mapped to `0+dataword'. This is similar to a standard Huffman compression scheme which focuses on compressing the most frequent dataword. However, as observed in Figure~\ref{fig:huffman}, even in the simplest Huffman code there are more than two different codeword lengths which is a significant drawback during hardware implementation of decoders. Our compression scheme can be built for any choice of $k$, and the resulting codeword length is either $1$ or $k+1$. 
Although we are increasing the number of 1s when we encode the all-0 dataword to `1' in the intermediate codeword, this is overcome by the subsequent mapping code.


Our compound code's rate depends on $k$ and the dataword statistics.  
As $k$ increases, both the maximum possible compression ratio for the all-0 dataword and the maximum possible rate for the other datawords increase. However, depending on the benchmark statistics, the frequency of the all-0 dataword varies at different granularities, and thus real-world results vary. Furthermore, because the compound codes contain a code with high energy savings, they also offer good energy savings.

\subsection{Summary of Promising LELCs}

We presented four classes of LELCs. Moreover, all of these LELCs have multiple viable design points. FnW can be applied at different granularities. Tree codes can be designed with many different topologies and assignments of datawords to leaves.  Mapping codes have many possible mappings and rates. Compound codes can exploit compression at different rates. 

In Table~\ref{tab:codes}, we list the ranges of rates and percentage energy reductions we experimentally observe across our CPU benchmarks for some of our most promising LELCs. The observed rate for a given benchmark is measured as the total number of dataword bits across all data payloads divided by the total number of codeword bits across all data payloads. For variable rate codes (below the dashed line in the table), a different rate is observed for each benchmark, because dataword statistics differ. The percentage energy reduction is measured by normalizing the total number of 1s in codeword bits across all data payloads by the total number of 1s in dataword bits across all data payloads. 
We include two different entries for each of our tree codes. TC1 and TC2
denote the codes as presented thus far, while TC1' and TC2' represent adjustments made to simplify their hardware implementations (see Section~\ref{sec:circuits}).
We also compare against Frequent Pattern Compression (FPC)~\cite{alaa}; as expected FPC improves rate, yet across all benchmarks, the energy consumed {\bf increases} due to an increase in the number of ones. This is consistent with the theoretical analysis in Sec.~\ref{sec:limit-study}.
This list of codes is not exhaustive, but it shows the potential of practical LELCs codes to reduce energy at differing rates.


\begin{table}[tb]
    \caption{Rates and percentage energy reduction for proposed LELCs observed across 8 PARSEC benchmarks running on a CPU. Codes above (below) the dashed line have fixed (variable) rate.
    \vspace{-0.05in}
    }
    \centering
    \small{
    \begin{tabular}{|c|c|c|}
    \hline
    Coding Scheme & Rate  & Energy Reduction \% \\
    \hline
    Flip-N-Write (k=3) & 0.75 & 15.52--23.66 \\
    Flip-N-Write (k=8) & 0.89 & 17.47--27.30 \\
    2-level FnW (k=4) & 0.76 &  24.31--36.40\\
    Mapping1 & 0.89 & 21.91--36.67\\
    Mapping2 & 1 &  10.79--30.41\\
    \hdashline
    TC1 & 0.76--0.78 &  18.30--27.88\\
    TC1' & 0.76 & 18.94--27.91\\ 
    TC2 & 0.92--0.97 &  11.10--20.83\\
    TC2' & 0.86--0.88 & 10.65--20.52\\ 
    Compound1 (k=32) &1.07--1.54& 15.90--25.74 \\
    \hline
    FP Compression~\cite{alaa} & 1.25--1.72 & -23.38 -- -10.74\\
         \hline
    \end{tabular}
    }
    \label{tab:codes} 
    \vspace{-0.1in}
\end{table}

\subsection{Dynamic Code Throttling}

Because many of our LELCs incur a cost in rate, they can potentially degrade performance.  When OCN link utilization is low (i.e., there is bandwidth slack in the OCN), a decrease in rate is unlikely to have much impact on performance.  However, when link utilization high, it can be more problematic to add to bandwidth demand.  
To limit the potential performance impact of coding, we explore disabling coding when OCN utilization is high.  (A more complicated solution would seek to switch between LELCs; we leave this option for future work.)
To implement throttling, we must answer two questions: (1) where to set the threshold and (2) how to measure OCN utilization.   
The choice of threshold is a trade-off.  If it is low (high), then coding is more (less) often disabled, so performance impacts are decreased (increased) but energy savings are also decreased (increased). We evaluate several different thresholds. 

Dynamic estimation of OCN utilization can be done in several ways, and we are agnostic as to how this is done.  The option that we use is a per-link scheme in which each router observes its local utilization at the granularity of $100$k cycles.  If the utilization is above a given threshold, coding is disabled for the subsequent $100$k cycles; otherwise, it is enabled. 
Local link utilization is tracked simply with a counter and a comparator, resulting in minimal circuit overheads. The choice of interval length is a tunable parameter, but we did not find that our results were very sensitive to it.

\section{Encoder/Decoder Circuitry}
\label{sec:circuits}

The hardware costs in terms of energy, latency, and area are critical to determining the practicality of our codes.
We implemented the circuits for all our LELC encoders and decoders in Verilog and synthesized them using Synposys Design Compiler with a $12$nm library. For codes requiring look-up tables, we use CACTI~\cite{cacti} with its smallest technology ($22$nm) to estimate the energy and latency, due to difficulties with the Synposys memory compiler. We discuss circuit implementation issues before presenting the synthesis results. 

\subsection{Fixed Rate Codes}

For our fixed-rate codes, we easily take advantage of parallelism.  Assume $d$-bit data payloads ($d=512$ in our experiments) and $k$-bit datawords. We can encode (decode) all $d/k$ datawords (codewords) in parallel, given enough replicas of the circuit that encodes (decodes) each $k$-bit dataword ($n$-bit codeword).  Replicas consume area and power (our power and area results in Sec.~\ref{sec:circuits} account for the overhead of these replicas), but the concurrency is vital for performance. 
For FnW, we have implemented circuits for various values of $k$.  For mapping codes, we have evaluated lookup tables (i.e., ROMs) for maps of different sizes. 

\subsection{Variable Rate Codes}

Variable-rate codes pose implementation challenges that we must overcome, because our two tree codes and our compound codes are variable-rate.  With a variable-rate code, parallelism is difficult because the boundaries between datawords and codewords may not be determined until runtime.
A variable-rate code can have variable-length datawords (like our tree codes) or codewords (like our compound codes),
both of which present implementation challenges.  If datawords (codewords) are variable-length, then we cannot start encoding (decoding) a dataword (codeword) until we have encoded (decoded) the previous one.  
Without parallelism, both encoding and decoding would require more latency than we can comfortably tolerate. 

\noindent\textbf{Parallel encoding with variable-length datawords.} We divide the data payload into fixed-length dataword chunks of size $f$ and use independent modules to process each dataword chunk in parallel. As the dataword lengths differ, a different number of datawords can fit into each fixed-length dataword chunk. We pack the datawords that fit into the dataword chunk and the remaining bits are sent unencoded. This solution lowers our energy reduction (because unencoded bits do not benefit from our coding), but may improve the rate of the overall scheme as unencoded bits have a rate of 1. This solution requires some circuit complexity to ``pack" the output of each module into flits, since the output length of each module is variable.  


\noindent\textbf{Padding variable-length codeword chunks into fixed-size codeword chunks.} Although tree codes have fixed-length codewords, parallelizing the encoding process with fixed-length dataword chunks (described above) results in variable-length codeword chunks. To overcome the complexity of parallel decoding of variable-length codeword chunks (see below), the encoder uses zero-padding to create fixed-length codeword chunks. Assume that $f$ dataword bits are encoded as at most $t$ codeword bits.  If any group of $f$ dataword bits are encoded to fewer than $t$ bits, the remaining bits can be ``padded" with 0s to get a fixed output length of $t$.  The padding hurts rate but has no impact on energy, which is attractive for tree codes that prioritize energy.
As a result of this technique, our implemented tree codes differ from the ones described earlier.
We divide data payloads into $32$-bit dataword chunks, and we use padding to create fixed-length codeword chunks of length $42$. We denote the more implementation-friendly versions of TC1 and TC2 as TC1' and TC2', respectively. 

\noindent\textbf{Decoding variable-length codewords.} Assume we receive a codeword string of $c$ bits.  If all codewords are the same length, $n$, or if we use zero-padding to create fixed-size codeword chunks, 
then parallel decoding is easy; we can have parallel modules to decode each group of $n$ codeword bits.
However, if codewords can be multiple possible lengths, as in our compound codes (without padding), then we do not immediately know where each codeword ends. Nevertheless, we can make the problem much more tractable by limiting the number of possible codeword lengths.  Specifically, if our codewords can be either of only two lengths $n_1$ or $n_2$, then the circuitry can, in parallel, consider all of the possible bit positions where codewords can start.  However, if codewords can have many different lengths, then the number of codeword starting positions is too large for efficient circuitry.
As such, our compound code has only two codeword lengths: one length for a run of 0s and a second length for all other data words.

Despite the availability of these three techniques for simplifying hardware implementations, TC2 remained too complicated and slow to be attractive. It requires padding for two different lengths to limit rate loss, and has more dataword lengths than TC1.  As TC1 is already our most complicated code to implement (see next section), we did not further pursue TC2.


\subsection{Latency, Energy, and Area Results}

The latency, energy consumption, and area footprints of our encoding and decoding circuits are summarized in Table~\ref{table:circuit}.
These circuits encode/decode 128 bits.
We target single-cycle latency for encoding/decoding to minimize performance impact.  All circuits fit within the latency of a 1.5 GHz clock period, and all but the encoder for TC1' fit within a 2GHz clock.
In terms of energy, to provide context, transmitting a {\it single bit} on a link consumes approximately $1-1.75$pJ~\cite{sun:nocs:2012, hotchips-2019}, whereas our encoding and decoding circuitry requires $1.5$pJ {\it per $128$-bit flit} for our compound code.
Each network interface will require an encoder and decoder; each encoder/decoder circuit processes 128-bits.
In terms of area, all our designs are negligible. 
 \begin{table}[tb]
 \caption{Latency/Energy/Area for Encoder/Decoder Circuits for 128-bit flits} 
 \label{table:circuit}
 \centering
 \small{
\begin{tabular}{|l|r|r|r|r|r|r|}
 \hline
  Coding & \multicolumn{2}{c|}{Latency (ps)} & \multicolumn{2}{c|}{Energy (pJ)} & \multicolumn{2}{c|}{Area ($\mu m^2$)}\\\cline{2-7}
   Scheme & Enc. & Dec. & Enc.  & Dec. & Enc. & Dec.\\
 \hline
 FnW (k=3) & 119.4 & 99.9 & 1.33 &  1.53 & 605 & 591\\
 FnW (k=8) & 203.1 & 104.6 & 1.27 &  1.33 & 669 & 515\\
 2L FnW (k=4) & 168.5 & 147.9 & 1.32 &  1.06 & 633 & 452\\
 Mapping1 & 74.9 & 85.7 & 0.32 &  0.65 & 682 & 1093 \\ 
 Mapping2 & 73.5 & 73.5 & 0.14 &  0.14 & 262 & 262 \\ 
 TC1' & 675.0 & 429.2 & 2.31 &  1.96 & 3775 & 2346 \\
 Comp1 (k=32) & 324.9 & 261.5 & 1.50 &  1.45 & 1226 & 1579 \\
 \hline
 \end{tabular}
 }
 \end{table}
\section{Experimental Methodology}
\label{sec:methodology}

By using line coding, we seek to reduce energy consumption, while minimizing the potential impact on performance and crosstalk.  To evaluate these metrics, we use full-system simulation and multithreaded benchmarks.
To model a multicore processor, we use the gem5 simulator~\cite{binkert:can:2011}.  We configure gem5 to model an x86 processor with 16 cores.   To model the OCN in detail, we use Garnet~\cite{agarwal:ispass:2009}. The configuration for our simulator can be found in Table~\ref{tab:simulator}. Performance results are produced directly by the simulator.  
The latencies in Table~\ref{table:circuit} are included in our simulation to properly account for encoding and decoding overheads. 
To estimate the energy savings for the OCN links,  we examine the bits in every message sent in the simulated system and code them accordingly. Energy is directly proportional to the number of 1s, and all energy results are normalized to the number of 1s in the uncoded baseline.


\begin{table}[tb]
    \caption{Simulation parameters}
    \centering
    \small{
    \begin{tabular}{|c|>{\centering\arraybackslash}m{5.0cm}|}
    \hline
         Cores & 16 OoO x86 cores  \\ 
         L1 ICache/DCache & I: 32KB, 2-way, D: 64KB, 2-way \\
         L2 & shared, inclusive 2MB, 8-way \\
         Cache block size & 64B \\
         Coherence protocol & MESI \\
         \hline
         Topology & 4$\times$4 mesh \\
         Routing Algorithm & minimal XY \\
         Router Latency & 1 cycle for uncoded baseline \\
         Link width & 16B (same as flit size) \\
         Flow Control & 4 virt. nets, each with 3 virt. channels \\
         \hline
    \end{tabular}
    \label{tab:simulator}
    }
\vspace{-0.1in}
\end{table}


We run applications from the  PARSEC~\cite{parsec} benchmark suite each with 16 threads, except for a few that do not currently run on our simulation infrastructure: \textit{dedup}, \textit{facesim}, \textit{raytrace}, \textit{vips}, and \textit{x264}.
Our use of benchmarks instead of assuming equiprobable dataword statistics distinguishes our work from much of the prior work. 
To highlight the broad applicability of our work, we also evaluate our LELCs on a discrete GPU using gem5's GCN3 GPU model~\cite{gutierrez:hpca:2018} and running 6 Pannotia graph workloads~\cite{che:iiswc:2013}.\footnote{Floyd-Warshall was omitted due to errors in the benchmark.} This system is configured with 32 CUs and uses a crossbar interconnect modeled with Garnet.


\section{Experimental Evaluation}
\label{sec:evaluation}

Our primary goal is to evaluate the potential of LELCs to reduce OCN energy without undue impact on performance.  We also evaluate the impact of LELCs on crosstalk.

\subsection{OCN Tolerance to Latency and Bandwidth}

    

\begin{figure}[t]
    
    \centering
    \includegraphics[trim={0in 0in 0in 0in},width=0.35\textwidth]{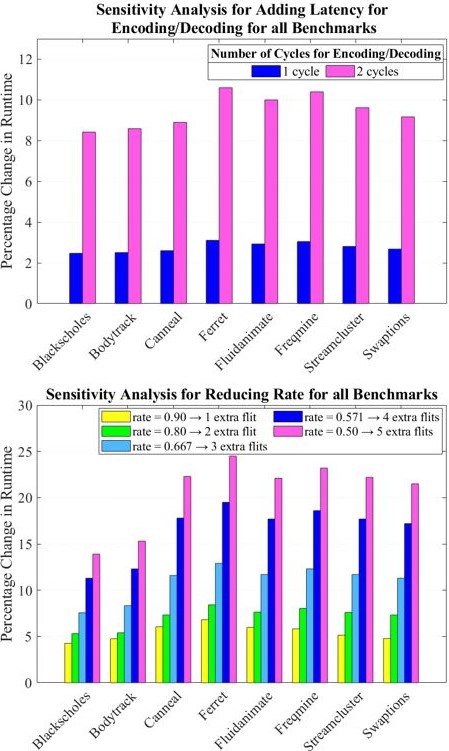}
    \vspace{-0.1in}
    \caption{Runtime impact of encoding/decoding latency (top) and code rate (bottom).} 
    \label{fig:sensitivity}
    \vspace{-0.2in}
\end{figure}

We first explore how much tolerance OCNs have to increases in latency (due to encoding/decoding time) and decreases in bandwidth (due to lower rate).

\subsubsection{Latency Tolerance of OCNs}

Encoding and decoding take some amount of time, and that added latency could impact OCN and overall system performance. To quantify this impact, we ran simulations in which we added varying amounts of latency for encoding and decoding.  The latency is paid once at sending (for encoding) and once at receiving (for decoding), but not at intermediate routers.  
The results of this sensitivity analysis, shown in Fig.~\ref{fig:sensitivity}, reveal that the overall performance impact of a 1-cycle latency is only $2\mhyphen3\%$, but at a 2-cycle latency the runtime increases by $8\mhyphen11\%$.\footnote{Since we do not encode the head flit, 1 cycle of encoding latency for body flits can be performed during the routing of the head flit. Our results, which assume a minimum 1 cycle encoding, are pessimistic in terms of performance as our simple circuitry can be fully overlapped with the head flit.}  
These results show that we are practically constrained to simple hardware that can complete within a single cycle.



\subsubsection{Bandwidth Tolerance of OCNs}

A rate reduction increases bandwidth demand. Modern OCNs are over-provisioned to tolerate bursty communication~\cite{fastpass,hesse2012fine} but in the common case have sufficient spare bandwidth to accommodate a rate reduction. 
Figure~\ref{fig:sensitivity} plots performance as a function of code rate.  We abstract away the codes themselves and simply apply a range of fixed rates, and we never throttle the coding.  
These results include the addition of one cycle each for encoding and decoding, as in the previous section.
The results show that adding one extra flit---which corresponds to a rate 
in the range $(0.8,1]$---has an impact of about $4\mhyphen7\%$ (including the roughly $2\mhyphen3\%$ for encoding/decoding latency).  Adding additional flits causes additional slowdown; unsurprisingly, doubling the number of flits per payload causes significant contention in the OCN and leads to excessive slowdowns.  These results motivate high-rate codes and code throttling. 



\subsection{Link Energy versus Performance}

\begin{figure*}[ht]
    \centering
    \includegraphics[clip=true, trim=0in 0in 0in 0in, width=0.98\textwidth]{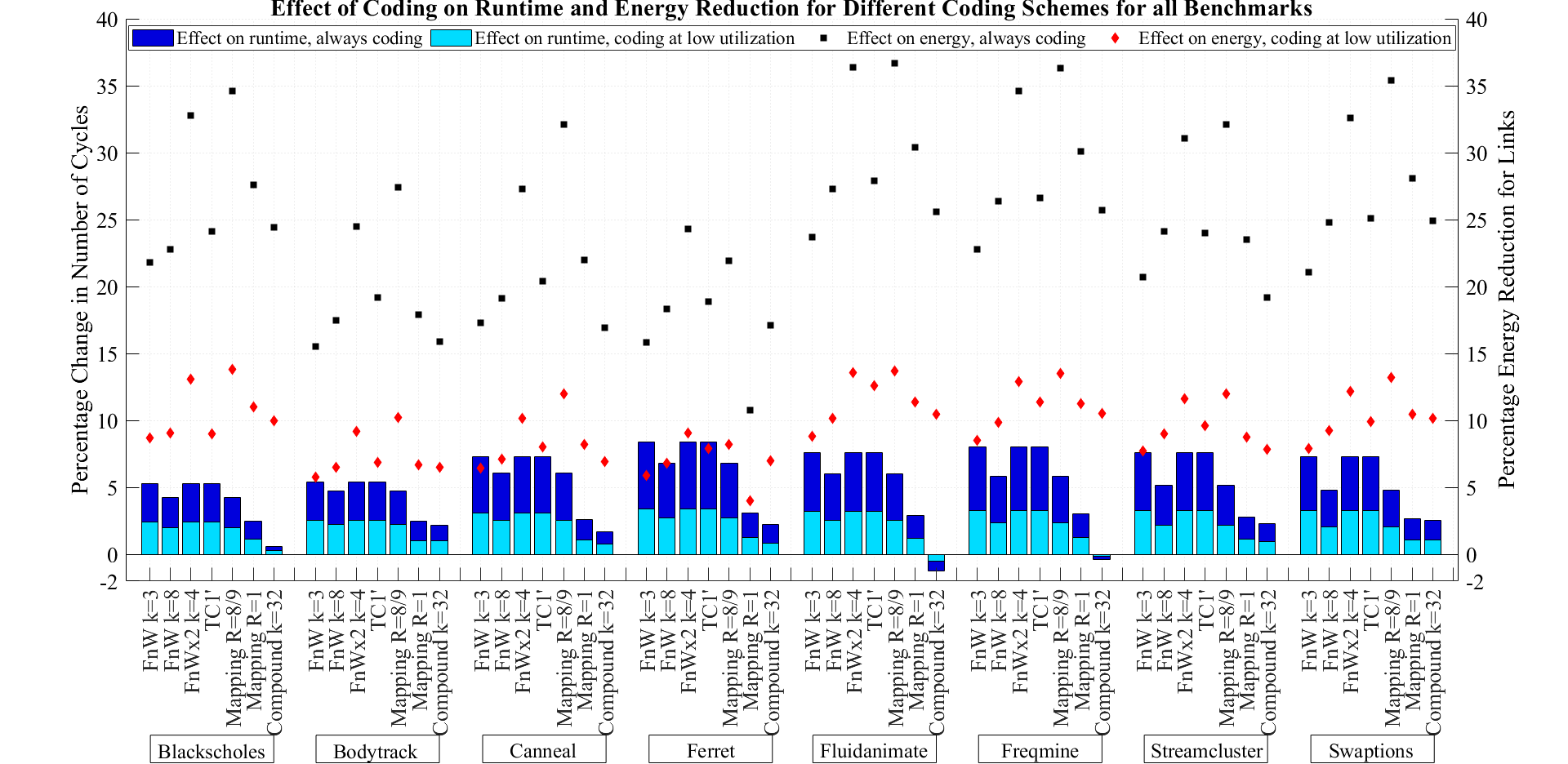}
    \caption{Percentage increase in runtime (left y-axis, bars) and percentage decrease in link energy (right y-axis, points)}
    \label{fig:perf-energy} 
    \vspace{-0.1in}
\end{figure*}

The fundamental trade-off with LELCs is link energy versus performance.  In Figure~\ref{fig:perf-energy}, we plot performance on the left y-axis and link energy on the right y-axis, with both represented as their percentage changes with respect to an uncoded baseline (increased for performance and decreased for energy consumption).  
Moreover, we plot results for both throttled and unthrottled coding. Energy consumption reductions are reports for OCN links; OCN links consume $\sim40\mhyphen50\%$ of the total energy in the network~\cite{kar2017case, werner2017designing, onur-asplos}.

\begin{figure}[t]
    \centering
    \includegraphics[clip=true, trim={1in 0in 0.4in 0.7in},width=0.44\textwidth]{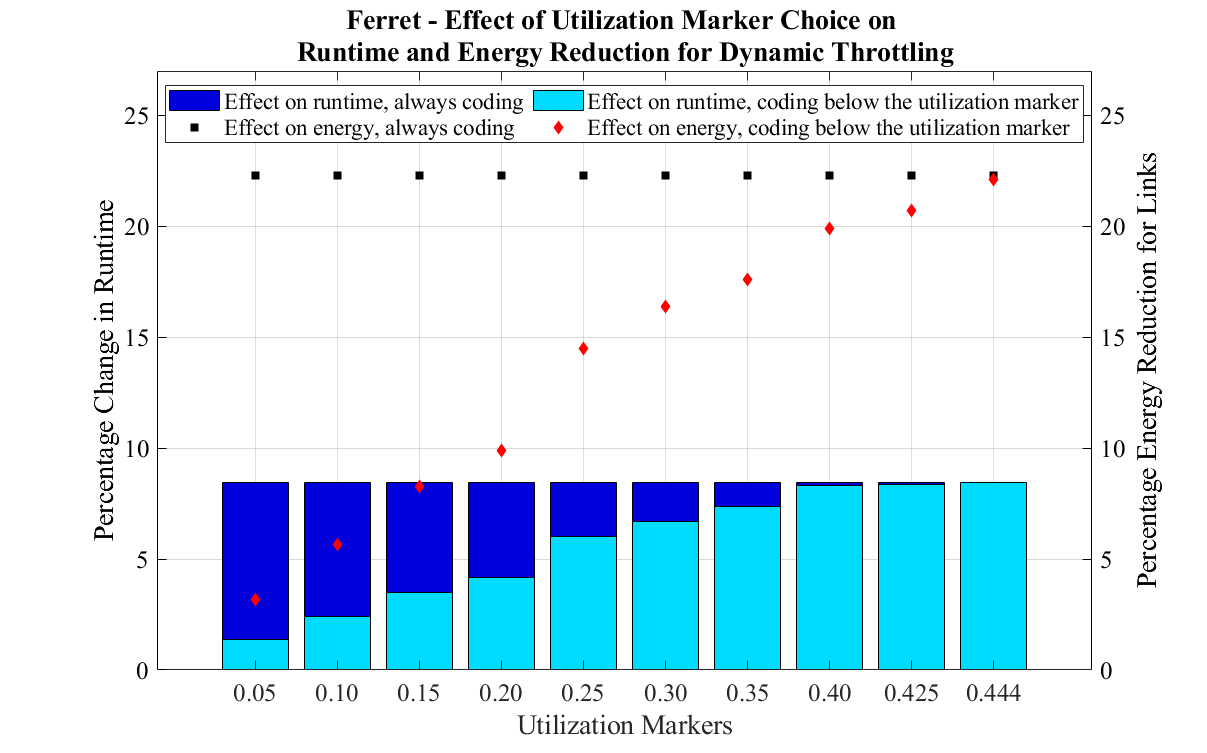}
    \vspace{-0.1in}
    \caption{Sweep of utilization markers for \textit{ferret}} 
    \label{fig:util_markers}
\vspace{-0.2in}
\end{figure}

The graph reveals three general insights. First, on every benchmark, every LELC provides a greater savings in energy than increase in runtime.  Second, the trends across benchmarks are consistent, just with modest differences in absolute values.  Third, dynamic throttling is effective in reducing performance impact; here, we throttle coding when utilization exceeds $16.5\%$ on a link (measured over $100$K cycles).  Dynamic throttling roughly halves the performance loss but at the cost of more than half of the energy savings. Figure~\ref{fig:util_markers} sweeps the utilization threshold values for \textit{ferret}. The utilization marker provides a knob to tune the performance-energy trade-off of our LELCs.

The graph also reveals trends across our LELCs:
\begin{itemize}
    \item Our rate-$8/9$ mapping code has the best energy reduction across benchmarks (except \textit{ferret}). 
    \item Our compound code has the least impact on runtime ($-1.25\mhyphen2.54\%$).  Due to its high rate, little performance degradation (and even small improvements for \textit{fluidanimate} and \textit{freqmine}) is observed. Compound codes benefit significantly from non-equiprobable application data; common, long strings of 0s are compressed for high rate.
    \item Our rate-$1$ mapping code achieves significant energy savings, with its only cost being the latency for encoding and decoding (i.e., there are no extra flits).  
    \item FnW$_{k=8}$ is pareto optimal with respect to FnW$_{k=3}$, because it offers greater energy savings and less runtime impact. It has lower runtime as it requires only 1 extra flit per data packet. 
    \item Our new 2-level FnW$_{k=4}$ has an almost identical rate as FnW$_{k=3}$---and they both incur the same number of additional flits, so they have the same performance---yet the 2-level scheme gets significantly better energy reduction. The same comparison is true for FnW$_{k=4}$ (not shown).
    \item 
    Our tree code (TC1') is designed for fair comparison with FnW$_{k=3}$.
    Designing a deeper tree, say to compete with FnW$_{k=8}$, would require complicated, slow circuitry. Compared to FnW$_{k=3}$, TC1 has lower rate, yet implementation-friendly TC1' offers greater energy savings with the same performance impact. 
    \item In this work, we assume one fixed code for each benchmark run; different codes may provide benefits for different phases of benchmark execution but come with added overhead of multiple encoders and decoders--we leave the exploration of this for future work.
    

\end{itemize} 
 
\begin{figure}[t]
    \centering
    \includegraphics[clip=true, trim=0in 0in 0in 0.37in, width=0.48\textwidth]{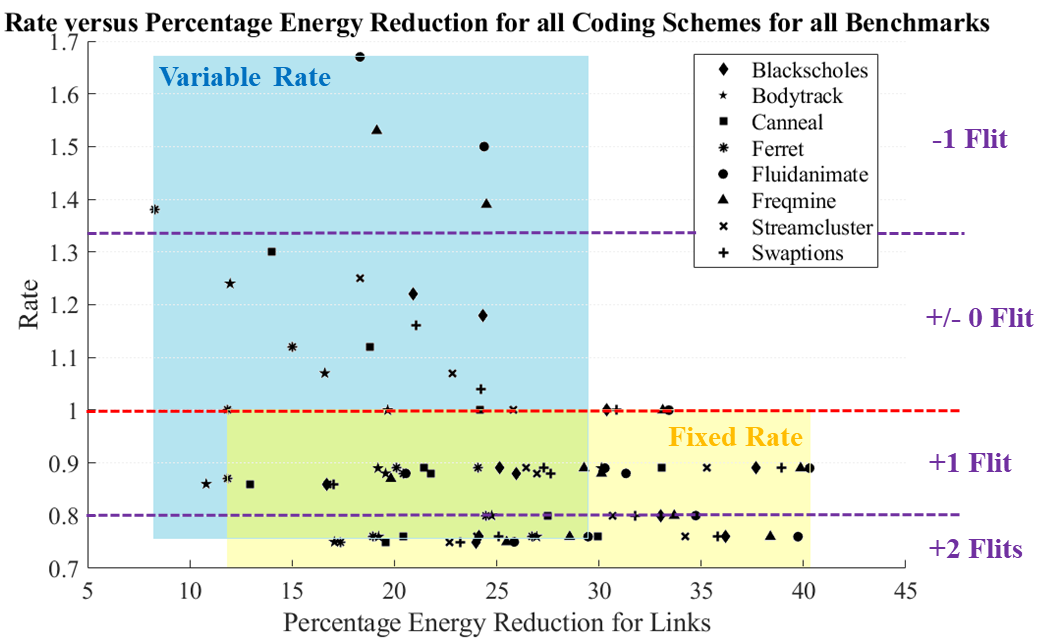}
    \caption{Rate vs. Energy}
    \label{fig:rate_energy_zones} 
\vspace{-0.2in}
\end{figure}

 To further understand the differences between LELCs, we plot rate (not end-to-end performance) versus link energy reduction in Figure~\ref{fig:rate_energy_zones}. 
 The figure includes all of the LELCs we have discussed and evaluated already, plus other variants (i.e., different values of $k$ for Flip-N-Write and compound codes, mapping codes with different maps, and different trees for tree codes). Including different variants for our codes provides insight into the size of the design space and the trade-offs that can be achieved. 
 In the figure, we shade zones based on whether the LELCs are fixed-rate or variable-rate. We observe that, in general, fixed-rate codes provide greater energy reductions, whereas variable-rate codes offer higher rates.

\noindent {\bf Variable-rate codes.} These LELCs, including implementation-friendly tree codes and compound codes, lie in the blue rectangle and provide a wide range of rates. The datapoints above rate-$1$ correspond to compound codes, with $k=16$ and $k=32$, that combine a simple compression scheme with our rate-$8/9$ mapping code. 
A deeper exploration of the importance of $k$ on compound codes reveals that $k<32$ offer higher rates, but less energy benefit and more complicated hardware (e.g., see Table~\ref{table:circuit} for $k=16$).
Choosing $k>32$ is unattractive because $32$ bits represents a common datatype, and we would not necessarily expect to have many runs of 0s at a granularity greater than $32$.  Thus, we consider $k=32$ to be the best choice for our compound codes. Variable-rate codes can also have rate less than $1$, shown by the shaded green (intersection) portion, which is where almost all of our tree codes lie. 


\noindent {\bf Fixed rate codes}. We explore a number of fixed-rate codes, which lie in the yellow rectangle and have rates of $1$,
$0.89$, $0.88$, $0.8$, $0.76$, and $0.75$.
For performance, we strongly prefer codes with rates above $0.8$, since they add only a single extra flit in each data packet.
In choosing codes between $0.8-1$, we prefer those with the greatest energy reduction (furthest to the right) and the simplest circuit implementation. Among the fixed-rate LELCs, 2-level FnW and the rate-$8/9$ mapping code offer the best trade-off between energy savings and implementation cost.

\subsection{Choosing a Map for Mapping Codes}

Our two mapping code results each use a single map for all benchmarks. This map was based on profiling all of the benchmarks; this is not optimal for any one specific benchmark. The profiles provide the 
frequency of all length-$8$ datawords, enabling us to map the most common datawords to the lowest-energy codewords (i.e,. the codewords with the fewest 1s).
 To explore the impact of using this aggregate map, rather than a per-benchmark map, we (a) profile each benchmark individually to generate a map that is optimal for that benchmark, (b) aggregate the per-benchmark profiles to create a single aggregate profile and single aggregate map, and (c) run every benchmark with every per-benchmark map and the aggregate map. 
Results (not shown due to space constraints) reveal that using the aggregate map performs well.  There is certainly a gap between using a benchmark's optimal map and using the aggregate map (2\%-8\%), but, even so, the mapping code with the aggregate map still achieves significant energy reductions.

\subsection{Impact on Crosstalk}

\begin{figure*}[ht]
    \centering
    \includegraphics[width=0.88\textwidth]{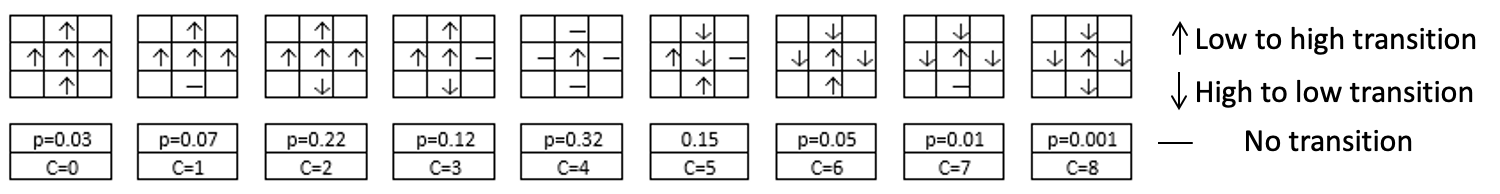}
    \caption{Crosstalk model from Patooghi et al.~\cite{patooghi:iwnoca:2019}. 
    There are $9$ pattern types; for each pattern we show one possibility but not the equivalent rotations or the ``mirror" version in which the low-to-high and high-to-low transitions are flipped.  Beneath each pattern we show the pattern's probability and its normalized crosstalk metric.} 
    \label{fig:crosstalk}
    \vspace{-0.1in}
\end{figure*}

With LELCs, we want to ensure that we do not increase crosstalk.  Since crosstalk is exacerbated by voltage transitions and LELCs reduce transitions, we expect LELCs to actually $reduce$ crosstalk.  
Many models have been developed for crosstalk, and we adopt the intuitive one from Patooghi et al.~\cite{patooghi:iwnoca:2019}.  
It models the crosstalk on a given {\it victim} wire in the middle of a 3$\times$3 grid as a function of its $8$ neighboring wires.  Neighboring wires that are horizontally or vertically adjacent have more impact than wires that are diagonally adjacent, because they are closer to the victim, and this particular model ignores the impact of diagonal neighbors. The amount of crosstalk depends heavily on the voltage transitions on the victim and its neighbors.  For example, if the victim is transitioning in the same direction (e.g., low-to-high) as its neighbors, that incurs less crosstalk than if they transition in opposite directions. The model, illustrated in Figure~\ref{fig:crosstalk}, measures crosstalk in terms of capacitance, normalized to the transition pattern with the least amount of capacitance. 

\begin{figure*}[ht]
\centering
\includegraphics[clip=true, trim=0 0 1.25in 0.36in, width=0.85\textwidth]{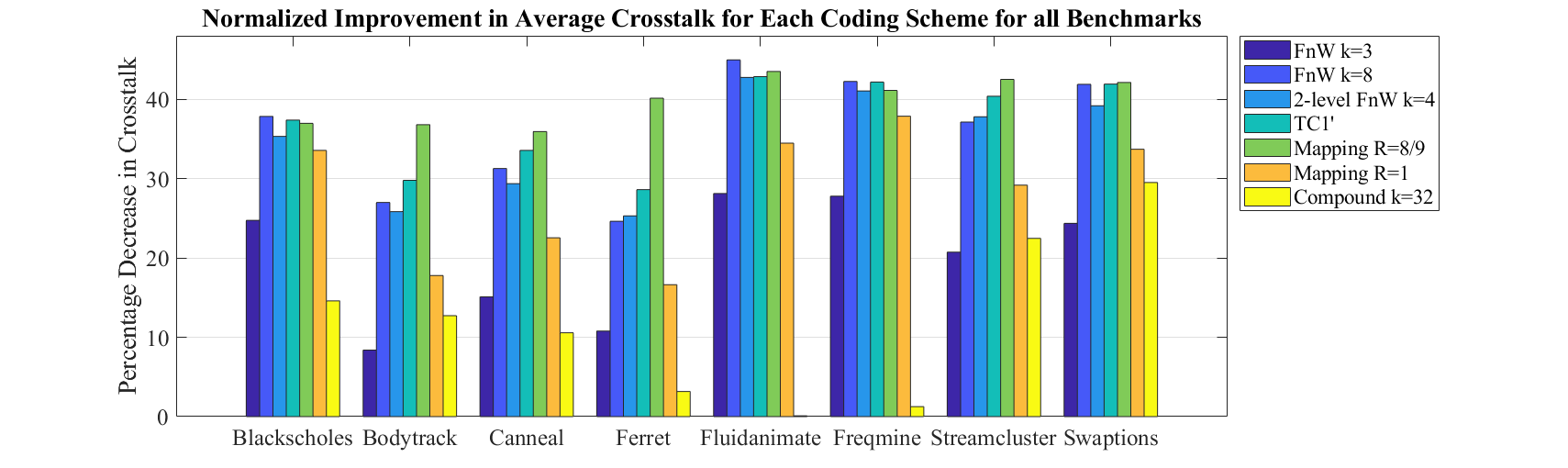}
\caption{The impact of LELCs on crosstalk} 
\label{fig:crosstalk-results} 
\vspace{-0.1in}
\end{figure*}

Using this model, we evaluate the crosstalk of each of our LELCs, normalized to that of an uncoded baseline.  The results in Figure~\ref{fig:crosstalk-results} show that, for every combination of LELC and benchmark, crosstalk is actually $less$ than the uncoded baseline.  This result is intuitive, because (a) our LELCs reduce the number of voltage transitions (represented by 1s in NRZI signalling), and (b) crosstalk effects are dominated by transitions (i.e., up or down arrows in Figure~\ref{fig:crosstalk}). 
The exact amount of crosstalk reduction, which ranges from negligible up to $45\%$, depends on the LELC and the benchmark.  Given that our primary concern was avoiding an $increase$ in crosstalk, this result is a bonus.
In addition, we observed that there is a future opportunity to optimize the mapping code used by the compound LELC to further reduce crosstalk.

\subsection{Recommended LELCs}

We recommend the use of Mapping1, 2-level FnW, and Compound1 LELCs.  These LELCs save energy up to $36.7\%$, $36.4\%$, and $25.7\%$, respectively, with less than $6.8\%$, $8.4\%$, and $2.5\%$ slowdown, respectively,
on our benchmarks.  The latency, energy, and area overheads for the encoder/decoder circuits are small for these LELCs (Table~\ref{table:circuit}). 



\section{Discussion}
\label{sec:discussion}


We have evaluated our LELCs in the context of general-purpose multicore architectures; however, they have broader applicability, some of which we briefly discuss here.

\noindent{\bf{GPUs.}} To assess the applicability of our codes beyond multi-core processors, we evaluated the link energy reduction for a GPU running Pannotia~\cite{che:iiswc:2013} applications. We present the results in Table~\ref{tab:gpu}. As observed in the Table, our LELCs designed for CPUs provide substantial opportunities for link energy savings in graph workloads running on a GPU.
In benchmarks such as $color$ and $mis$, we note very high gains for some of our codes. 
We discovered that some benchmarks initialize matrices to -1 (0xFFFFFFFF), which can be effectively coded for high gains. 
Another property we see is matrices initialized to very large positive values that also feature a large number of 1s. These results highlight a few key findings:
\begin{itemize}
\item Common programmer behavior lends itself to non-equiprobable data; properly designed codes can transform this data into more efficient values to be transmitted.
\item Our codes, which were designed based on CPU workloads, still perform very well on GPU workloads, even though dataword characteristics may differ. 
\item One GPU benchmark (SSSP) shows lower gains--this benchmark is dominated by zeros, leaving little opportunity for coding. The remaining GPU benchmarks out-perform the CPU benchmarks in terms of energy reduction. 
\end{itemize}

\begin{table}[tb]
    \caption{Rates and percentage energy reduction for proposed LELCs observed across 6 Pannotia benchmarks running on a GPU. Codes above (below) the dashed line have fixed (variable) rate.
    }
    \centering
    \small{
    \begin{tabular}{|c|c|c|}
    \hline
    Coding Scheme & Rate  & Energy Reduction \% \\
    \hline
    Flip-N-Write (k=3) & 0.75 & 9.57--50.23 \\
    Flip-N-Write (k=8) & 0.89 &  11.22--67.69\\
    2-level FnW (k=4) & 0.76 & 15.79--73.39\\
    Mapping1 & 0.89 & 19.24--68.75\\
    Mapping2 & 1 & 7.38--65.53\\
    \hdashline
    TC1 & 0.77--0.88 & 10.75--59.50 \\
    TC1' & 0.76 & 11.89--55.14\\ 
    TC2 & 0.96--1.06 & 5.62--49.28 \\
    TC2' & 0.89 & 5.66--41.53\\ 
    Compound1 (k=32) & 0.87--1.51 & 10.11--60.49 \\
         \hline
    \end{tabular}
    }
    \label{tab:gpu} 
    \vspace{-0.1in}
\end{table}

\noindent {\bf Chiplet-based Designs.} 
Chiplet-based architectures have emerged as a means to combat skyrocketing manufacturing costs by manufacturing smaller dies and composing them into larger systems through interposers or package-level interconnects~\cite{amd-isca}. 
Links between chiplets consume more energy than on-chip links; current efficiencies on inter-chiplet links are around 2pJ per bit~\cite{amd-isca, hotchips-2019}. 
Thus we can expect more benefit from our LELCs in this context. No modifications are needed to our design to support chiplets---inter-chiplet routers code the data prior to sending and decode upon receipt.

\noindent {\bf Machine-learning accelerators.}
Prior work has observed that the distribution of activation and weight values in deep convolutional neural networks are non-uniform~\cite{cnvultin,additive}.
The non-uniform nature of values in ML models would make their communication highly-amenable to the types of LELCs explored in this paper.

\section{Related Work}
\label{sec:related-work}






OCN line coding has been studied to reduce power and crosstalk. A temporal coding scheme codes the current flit with the flit transmitted ahead of it on the link~\cite{palesi:tcad2011,palesi:euromicro2009} using a bus-invert style scheme; this works well for wormhole routed OCNs where flits from different packets cannot be interleaved. However, in modern OCNs, virtual channel flow control prevails, which can disrupt the flow of flits from the same packet, rendering such techniques ineffective.
Other work focuses on uniform random data~\cite{jantsch:iscas2005} which leaves opportunity on the table, as we demonstrate on benchmark data; that work also assumed hop-by-hop encoding which consumes more overhead. 
Finally, an XOR-coding scheme improves the efficiency of speculative routers~\cite{nox}.


Coding research in OCNs has largely applied techniques from DRAM bus coding, which has a longer history of exploration.
Modern DRAMs often employ Data Bus Inversion (DBI) based on Bus-Invert coding~\cite{stan:1995} to reduce energy and noise. DBI is analogous to FnW discussed earlier in the paper. 
A bus coding scheme that exploits value locality has been proposed~\cite{bishop2001}.
Bitwise Difference Encoding~\cite{Seol:isca2016} exploits data similarity to reduce the Hamming weight of words on the DRAM bus. 
Multiple schemes that XOR adjacent bits to reduce transitions on a link have been proposed~\cite{lee:iccad2004,lee:hpca2018}
An online clustering method that XORs data with a common value to reduce the number of 1s transmitted improves on DBI~\cite{wang2016reducing}; this scheme requires multiple cycles for encoding and would have high overhead to synchronize the center values across the OCN.



Primarily in the CAD/EDA community, there has been research on crosstalk-avoidance codes for OCNs and buses~\cite{duan:hot-interconnects:2001, kumar:date:2013, shirmohammadi:isrccsc:2015, soleimani:iccd:2017, sridhara:vlsi2005, zou:aspdac:2014}.
This prior work develops models (like the one we use from this paper~\cite{patooghi:iwnoca:2019}) that are a function of the geometry of the wires and the signals that are on each wire.  Certain signal patterns (e.g., a wire transitioning from low-to-high while its neighbors transition from high-to-low) 
are worse than others, and the crosstalk-avoidance codes seek to eliminate the worst-case patterns of signals.  
This prior work also assumes that inputs are equiprobable, which enables mathematical analysis but is not representative of real-world inputs. Unlike prior work, we reduce crosstalk by changing the distribution of signal patterns to favor patterns with fewer 1s, rather than by eliminating a few select patterns.

Compression schemes are similar but distinct from coding.
Compression has been studied in the OCN~\cite{das:hpca2008}, including table-based compression~\cite{jin:MICRO2008}, delta-compression~\cite{noc-delta}, and approximate (lossy) compression~\cite{boyapati:isca2017}.  
As noted earlier, coding is theoretically superior to compression so that is our focus; in addition, we demonstrate that coding provides greater benefits when compared to frequent pattern compression~\cite{alaa}. 
Frequent pattern compression requires multiple clock cycles, whereas our codes can be implemented within a single cycle.

\section{Conclusions}
\label{sec:conclusions}

Communication in various forms consumes substantial on-chip energy. Fundamentally rethinking the representation of data can reduce energy spent on communications. Line coding originating in the information theory community offers insight into more efficient on-chip communication. In this paper, we apply line coding to OCN communication.
Two particular insights are gained from our results: 1) designing codes based on equiprobable data does not achieve the full gains possible in real applications and 2) the discrete flit sizes in OCNs impact the desirable line code rates. Furthermore, our line codes have the added benefit of reducing crosstalk.  

\section*{Acknowledgments}

This material is based on work supported by the National Science Foundation under grant CCF-171-7602. Any opinions, findings and conclusions or recommendations expressed in this material are those of the authors and do not necessarily reflect the views of the U.S. National Science Foundation.
We gratefully acknowledge the support of the Natural Sciences and Engineering Research Council of Canada (NSERC) Discovery Grant RGPIN-2020-04179. This research was undertaken, in part, thanks to funding from the Canada Research Chairs program and through the support of the University of Toronto McLean Award.

\balance

\bibliographystyle{IEEEtranS}
\bibliography{refs}

\end{document}